\documentclass[lettersize,journal]{IEEEtran}

\usepackage{cite}

\usepackage{algorithm}
\usepackage{algorithmic}
\usepackage{graphicx}
\usepackage{subcaption}
\usepackage{textcomp}
\usepackage{xcolor}
\usepackage{multirow}
\usepackage{dirtytalk}
\usepackage{hyperref}
\usepackage{balance}
\usepackage{comment}
\usepackage{amsmath,amssymb,amsfonts,mathtools,bm}

\makeatletter
\@ifundefined{lemma}{}{}
\@ifundefined{theorem}{}{}
\@ifundefined{proposition}{\newtheorem{proposition}{Proposition}}{}
\@ifundefined{corollary}{}{}
\makeatother

\newcommand{\CN}{\mathcal{CN}}
\newcommand{\E}{\mathbb{E}}

\newcommand{\jj}{\mathrm{j}}
\newcommand{\indep}{\mathrel{\perp\mspace{-10mu}\perp}}

\usepackage[figurename=Fig.,labelsep=period,font=small]{caption}

\newcommand{\orcidauthorA}{\href{https://orcid.org/0000-0001-5792-0842}{\includegraphics[scale=0.05]{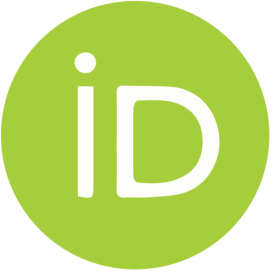}}}
\newcommand{\orcidauthorB}{\href{https://orcid.org/0000-0002-1478-2272}{\includegraphics[scale=0.05]{Letter_figures/Shah_WCL_fig_orcid.eps}}}
\newcommand{\orcidauthorC}{\href{https://orcid.org/0000-0003-4363-396X}{\includegraphics[scale=0.05]{Letter_figures/Shah_WCL_fig_orcid.eps}}}

\begin{document}

\title{Low-Overhead CSI Prediction via Gaussian Process Regression}	
	\author{{Syed Luqman Shah$^{\orcidauthorA{}}$, \textit{Graduate Student Member}, \textit{IEEE},  Nurul Huda Mahmood$^{\orcidauthorB{}}$, \textit{Member}, \textit{IEEE}, \\ and Italo Atzeni$^{\orcidauthorC{}}$, \textit{Senior Member}, \textit{IEEE}
    }\\
    \thanks{The authors are with Centre for Wireless Communications, University of Oulu, Finland (e-mail: \{syed.luqman, nurulhuda.mahmood, italo.atzeni\}@oulu.fi). This work was supported by the Research Council of Finland (336449 Profi6, 348396 HIGH-6G, 359850 6G-ConCoRSe, and 369116 \href{https://www.6gflagship.com/}{6G Flagship}). Reproducible code for this work is available at:~\url{https://github.com/Syed-Luqman-Shah-19/MIMOGPR1}.}}


\maketitle
\begin{abstract}
Accurate channel state information (CSI) is critical for current and next-generation multi-antenna systems. Yet conventional pilot-based estimators incur prohibitive overhead as antenna counts grow. In this paper, we address this challenge by developing a novel framework based on Gaussian process regression (GPR) that predicts full CSI from only a few observed entries, thereby reducing pilot overhead. The correlation between data points in GPR is defined by the covariance function, known as kernel. In the proposed GPR-based CSI estimation framework, we incorporate three kernels, i.e., radial basis function, Mat\'ern, and rational quadratic, to model smooth and multi-scale spatial correlations derived from the antenna array geometry. The proposed approach is evaluated across two channel models with three distinct pilot probing schemes. Results show that the proposed GPR with 50\% pilot saving achieves the lowest prediction error, the highest empirical 95\% credible-interval coverage, and the best preservation of spectral efficiency relative to the benchmarks. 
\end{abstract}

\section{Introduction}
\IEEEPARstart{M}{assive} multiple‐input multiple‐output (MIMO) systems are expected to play a pivotal role in 6G wireless networks, but their promised gains rely on accurate channel state information (CSI)~\cite{Nandana_white}. The pilot overhead required by classical estimators, such as least squares (LS) and minimum mean squared error (MMSE), scales unfavorably with the number of transmit antennas~\cite{Matrix_Completion_Zhiwei2025}. Compressed sensing (CS) techniques reduce this burden by exploiting channel sparsity, though they suffer from power leakage due to off-grid points~\cite{Matrix_Completion_Zhiwei2025, Signal_Recovery_2010}. For instance, low-rank matrix completion mitigates overhead through methods like nuclear-norm minimization or the alternating direction method of multipliers~\cite{ADMM_algo_Sig_letter_2018}, including extensions with subgradient descent and joint nuclear- and $\ell_1$-norm formulations~\cite{Matrix_Completion_Zhiwei2025}. However, these approaches are sensitive to rank errors, noise, and violations of low-rank assumptions in rich-scattering environments.

In contrast, Gaussian process regression (GPR) is a nonparametric, Bayesian estimator that provides calibrated uncertainty~\cite{shah2025}. The covariance function provides domain knowledge, so GPR is agnostic to sparsity and bandlimitedness~\cite{ADMM_algo_Sig_letter_2018}. GPR exploits joint transmit-receive correlations and captures the nonseparable structure observed in measured channels~\cite{Jieao2025IEEE_TIT}. Its performance does not depend on the restricted isometry property or low coherence~\cite{Signal_Recovery_2010}, and it remains effective with equispaced or hardware-constrained activations~\cite{Arj25}. Unlike CS, GPR requires only a second-order description, making it resilient to off-grid leakage, diffuse multipath, and rank/sparsity variability~\cite{Matrix_Completion_Zhiwei2025}. It also returns per-entry posterior uncertainty, enabling principled pilot budgeting, reliability targeting, and link adaptation: these capabilities are absent in interpolation-based and standard CS solvers. Recent works have used GPR, e.g., to incorporate electromagnetic priors into statistical channel models~\cite{Jieao2025IEEE_TIT}, predict interference power for proactive 6G resource allocation~\cite{shah2025}, design spatio-temporal kernels grounded in electromagnetic theory to improve CSI prediction~\cite{li2024spatioGPR}, and model distortion induced by hardware impairments to enhance channel estimation~\cite{Arj25}. Taken together, GPR learns spatial correlations from limited observations and quantifies uncertainty, enabling informed pilot placement and reconstruction confidence.

This work addresses spatial interpolation in MIMO systems with limited pilot measurements. We formulate the CSI prediction as a Gaussian process (GP) inference, enabling full channel reconstruction with quantified uncertainty. A GPR framework is developed to reduce pilot overhead by estimating the complete CSI from a small subset of pilot observations. The correlation between data points in GPR are defined by the covariance function, known as kernel, which captures the smooth and multi-scale spatial correlations inherent in wireless channels. We investigate three geometric kernels in our proposed GPR framework, i.e., radial basis function (RBF), Mat\'ern, and rational quadratic (RQ). As an additional contribution, we propose a novel antenna probing design to reduce the pilot length in each coherence interval. 

Our proposed approach is validated on the Kronecker~\cite{kronecker2002} and Weichselberger~\cite{weichselberger2006} channel models, with three reduced-pilot antenna-probing schemes assessed for each. Although these channel models naturally align with the GP prior, we prove that the resulting GP posterior mean is equivalent to the best linear unbiased predictor (BLUP)~\cite{BLUP}, without assuming channel Gaussianity. Simulation results under both models demonstrate that our proposed framework can achieve lower prediction error, higher empirical $95\%$ credible-interval coverage across all kernels, and the closest spectral efficiency (SE) corresponding to the true channel compared to conventional LS and MMSE estimators while enabling up to $50\%$ reduction in pilot overhead without compromising the SE.

\textit{\textbf{Notations:}} Bold lowercase and uppercase letters denote vectors and matrices, respectively. $\mathbf{I}_P$ denotes the $P\times P$ identity matrix and $\mathbf{e}_i$ is the $i$th canonical basis vector. $\mathrm{vec}(\cdot)$ denotes vectorization. $(\cdot)^{\mathsf{H}}$ denotes the Hermitian transpose. $\lVert\cdot\rVert_{\mathrm{F}}$ denotes the Frobenius norm. $\mathrm{tr}(\cdot)$ denotes the trace and $\det(\cdot)$ represents the determinant. $\mathrm{diag}(\cdot)$ converts a diagonal matrix to a vector and a vector to a diagonal matrix. $\otimes$ denotes the Kronecker product. $\CN(\cdot,\cdot)$ denotes a circularly symmetric complex Gaussian distribution. $k_{\boldsymbol{\theta}}(\cdot,\cdot)$ is a kernel function and $\mathbf{K}$ is the corresponding kernel matrix. $\gamma$, $\ell$, $\alpha$, and $\nu$ denote the kernel’s scaling, lengthscale, exponential-decay, and smoothness parameters, respectively (model-specific where applicable). $\E[\cdot]$ denotes expectation.

\section{System Model}
We consider a narrowband point-to-point MIMO link with $N_{\textrm{t}}$ transmit and $N_{\textrm{r}}$ receive antennas.
During training, $n_{\textrm{t}}$ of the $N_{\textrm{t}}$ transmit antennas are active and transmit the pilot matrix
$\mathbf{S}\in\mathbb{C}^{n_{\textrm{t}}\times T}$, whose rows are orthonormal, i.e., $\mathbf{S}\mathbf{S}^{\mathsf{H}}=\mathbf{I}_{n_{\textrm{t}}}$, which implies $T\ge n_{\textrm{t}}$. Let $\mathbf{F}\in\{0,1\}^{N_{\textrm{t}}\times n_{\textrm{t}}}$ be the antenna selection matrix satisfying $\mathbf{F}^{\mathsf{H}}\mathbf{F}=\mathbf{I}_{n_{\textrm{t}}}$. The transmitted signal is $\mathbf{X}=\mathbf{F}\mathbf{S}\in\mathbb{C}^{N_{\textrm{t}}\times T}$ and the received signal is given by $\mathbf{Y}=\mathbf{H}\mathbf{X}+\mathbf{N}\in\mathbb{C}^{N_{\textrm{r}}\times T}$, where $\mathbf{H}\in\mathbb{C}^{N_{\textrm{r}}\times N_{\textrm{t}}}$ is the channel matrix (with $(i,j)$th entry $H_{i,j}$) and $\mathbf{N}\in\mathbb{C}^{N_{\textrm{r}}\times T}$ is the noise matrix with independent $\CN(0,\sigma_\textrm{n}^2)$ entries.
After decorrelation via right-multiplication by $\mathbf{S}^{\mathsf{H}}$, we obtain $\tilde{\mathbf{Y}}=\mathbf{Y}\mathbf{S}^{\mathsf{H}}=\mathbf{H}\mathbf{F}+\mathbf{N}\mathbf{S}^{\mathsf{H}}\in\mathbb{C}^{N_{\textrm{r}}\times n_{\textrm{t}}}$, whose columns correspond to the observed channel vectors associated with the $n_{\textrm{t}}$ transmit antennas.

We adopt both the Kronecker~\cite{kronecker2002} and Weichselberger~\cite{weichselberger2006} channel models. Under the Kronecker model, we have $\mathbf{H}=\mathbf{R}_{\textrm{r}}^{1/2}\mathbf{G}\mathbf{R}_{\textrm{t}}^{1/2}$, where $\mathbf{R}_{\textrm{t}}\in\mathbb{C}^{N_{\textrm{t}}\times N_{\textrm{t}}}$ and $\mathbf{R}_{\textrm{r}}\in\mathbb{C}^{N_{\textrm{r}}\times N_{\textrm{r}}}$ are the transmit and receive covariance matrices, and $\mathbf{G}\in\mathbb{C}^{N_{\textrm{r}}\times N_{\textrm{t}}}$ has independent $\CN(0,1)$ entries. 
Under the Weichselberger model, we have $
\mathbf{H}=\mathbf{U}_{\textrm{r}}\mathbf{\tilde{A} }\mathbf{U}_{\textrm{t}}^{\mathsf{H}}$, 
where $\mathbf{U}_{\textrm{t}}$ and $\mathbf{U}_{\textrm{r}}$ are unitary matrices from the eigen decompositions 
$\mathbf{R}_{\textrm{t}}=\mathbf{U}_{\textrm{t}}\boldsymbol{\Lambda}_{\textrm{t}}\mathbf{U}_{\textrm{t}}^{\mathsf{H}}$ and 
$\mathbf{R}_{\textrm{r}}=\mathbf{U}_{\textrm{r}}\boldsymbol{\Lambda}_{\textrm{r}}\mathbf{U}_{\textrm{r}}^{\mathsf{H}}$, respectively. The matrix $\mathbf{\tilde{A}}\in\mathbb{C}^{N_{\textrm{r}}\times N_{\textrm{t}}}$ has independent entries $\tilde{A}_{ij}\sim\CN(0,\Omega_{ij})$, where $\Omega_{ij} \in \mathbb{R}_+$ represents the power coupling coefficient between the $i$th receive and $j$th transmit eigenmodes. Collecting the coefficients $\{\Omega_{ij}\}$ into the matrix $\boldsymbol{\Omega} \in \mathbb{R}_+^{N_{\textrm{r}} \times N_{\textrm{t}}}$, we have $\mathbf{R}_{\textrm{H}} = (\mathbf{U}_{\textrm{t}}^{*}\otimes \mathbf{U}_{\textrm{r}}) \mathrm{diag} \big(\mathrm{vec}(\boldsymbol{\Omega})\big) (\mathbf{U}_{\textrm{t}}^{\mathsf{T}}\otimes \mathbf{U}_{\textrm{r}}^{\mathsf{H}})$.

\begin{figure}[t]
\centering
\includegraphics[width=\linewidth]{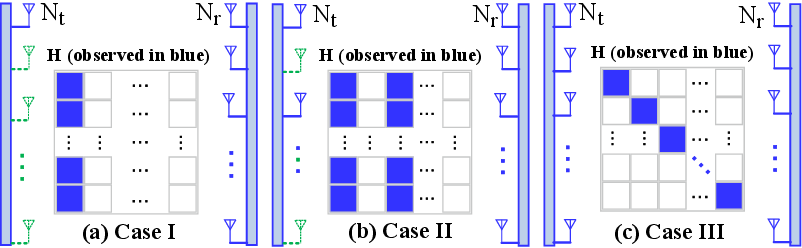}
\caption{Illustrative geometries used for the evaluation of the proposed framework, with $n_{\textrm{t}}$ active transmit antennas shown in blue; inactive (green) antennas do not transmit pilots. (a) Case~I: $n_{\textrm{t}} = 1$ active transmit antenna; (b) Case~II: $n_{\textrm{t}} = \lceil N_{\textrm{t}}/2\rceil$ active transmit antennas; (c) Case~III: $n_{\textrm{t}} =\min(N_{\textrm{t}},N_{\textrm{r}})$ active transmit antennas with one-to-one transmit-receive antenna pairing.}

\label{fig:System_Model}
\end{figure}

To relax the constraint on the pilot length from $T \ge N_{\textrm{t}}$ to $T \ge n_{\textrm{t}}$ within a coherence block, we introduce an antenna-probing design in which the transmitter activates the antennas in the set $\mathcal A\subseteq\{1,\ldots,N_{\textrm t}\}$, with $n_{\textrm t}=|\mathcal A|$, while all the $N_{\textrm r}$ receive antennas listen. The corresponding columns of $\mathbf{H}$ are observed and serve as training data for GPR, which infers the remaining entries of $\mathbf{H}$ as described next in Section~\ref{sec:GPR}. The proposed GPR-based predictor applies to any probing scheme; here we evaluate it under three illustrative geometries:
\begin{itemize}

\item \noindent {\emph{Case~I:} only the first transmit antenna is active, i.e., $n_{\textrm{t}} =1$, so we observe the first channel column $\mathbf{H}(:,1)$ using all the receive antennas (see Fig.~\ref{fig:System_Model}(a)). This serves as a stress test (failure mode).}

\item \noindent {\emph{Case~II:} we activate $n_{\textrm{t}}=\lceil  N_{\textrm{t}}/2\rceil$ equispaced transmit antennas, forming the set $
\mathcal A(n_\textrm{t})=\{1+\lfloor(m-1)\tfrac{N_\textrm{t}-1}{n_\textrm{t}-1}\rfloor: m=1,\dots,n_\textrm{t}\}$,
and observe the channel columns $\mathbf{H}(:,\mathcal{A}(n_{\textrm{t}}))$ (see Fig.~\ref{fig:System_Model}(b)). This design offers a favorable pilot-efficiency trade-off.}

\item \noindent {\emph{Case~III:} a diagonal-anchored scheme that activates $n_{\textrm{t}} =\min(N_{\textrm{t}},N_{\textrm{r}})$ transmit antennas as in full training, but retains only the diagonal channel entries $\{H_{i,i}\}$ for training. This emulates one-to-one transmit-receive pairing while discarding all the off-diagonal samples (see Fig.~\ref{fig:System_Model}(c)). This is not a pilot-saving design; rather, it isolates sampling-geometry effects under the same feasibility condition $T\ge n_{\textrm t}$.}
\end{itemize}

\section{GPR for Channel Prediction}
\label{sec:GPR}

\begingroup

\begin{algorithm}[t]
\caption{{GPR-Based MIMO Channel Prediction}}
\label{alg:gpr_channel_prediction}
\begin{algorithmic}[1]
\REQUIRE Training points $\mathbf{Z}\in\mathbb N^{P\times2}$, observations $\mathbf{h}\in\mathbb C^P$, test points $\mathbf{Z}_*\in\mathbb N^{M\times2}$, $k_{\boldsymbol{\theta}}(\cdot,\cdot)$, initial $\boldsymbol{\theta}$
\ENSURE Predicted complex entries $\mathbf{h}_*\in\mathbb C^M$ and posterior covariances $\boldsymbol{\Sigma}_*^{\mathrm{Re}/\mathrm{Im}}\in\mathbb R^{M\times M}$
\STATE Split the observations, i.e., $\mathbf{h}^{\mathrm{Re}}=\mathrm{Re}(\mathbf{h})$, $\mathbf{h}^{\mathrm{Im}}=\mathrm{Im}(\mathbf{h})$
\STATE Maximize $\log p(\mathbf{h}^{\mathrm{Re}}| \mathbf{Z},\boldsymbol{\theta})+\log p(\mathbf{h}^{\mathrm{Im}}| \mathbf{Z},\boldsymbol{\theta})$ to get $\boldsymbol{\theta}^\star$
\STATE Build $\mathbf{K}$, $\mathbf{K}_*$, and $\mathbf{K}_{**}$ from $k_{\boldsymbol{\theta}}(\cdot,\cdot)$ and $\boldsymbol{\theta}^\star$
\STATE Compute the posteriors $\boldsymbol{\mu}_*^{\mathrm{Re}/\mathrm{Im}}$ and $\boldsymbol{\Sigma}_*^{\mathrm{Re}/\mathrm{Im}}$
\STATE Form complex prediction: $\mathbf{H}_* = \boldsymbol{\mu}_*^{\mathrm{Re}} + \jj\,\boldsymbol{\mu}_*^{\mathrm{Im}}$
\STATE \textbf{return} $\mathbf{H}_*$ and $\boldsymbol{\Sigma}_*^{\mathrm{Re}/\mathrm{Im}}$
\end{algorithmic}
\end{algorithm}
\endgroup

Define the antenna grid $\mathcal{G}=\big\{ (i,j) : i \in \{ 1, \ldots, N_{\textrm{r}} \}, j \in \{ 1, \ldots, N_{\textrm{t}} \} \big\} \subset \mathbb{N}^{2}$. We model the channel entries as a GP over $\mathcal{G}$ and use GPR to predict the missing entries from the observed ones via kernel-based spatial correlations. This section outlines the training set and prediction, kernel selection, hyperparameter optimization, and complex channel reconstruction. The procedure is summarized in Algorithm~\ref{alg:gpr_channel_prediction}.

 \textit{\textbf{Training set and prediction:}} Let $\mathcal{X} \subset \mathcal{G}$ be the set of observed indices (corresponding to the training points), with $|\mathcal{X}| = P$. Moreover, let $\mathcal{X}_{*} = \mathcal{G} \setminus \mathcal{X}$ be the set of indices to be predicted (corresponding to the test points), with $|\mathcal{X}_{*}| = M$. Define the matrix of training points $\mathbf{Z} = [\mathbf{z}_{1}^{\mathsf T}, \ldots, \mathbf{z}_{P}^{\mathsf T}]^{\mathsf T} \in\mathbb{N}^{P \times 2}$, with $\mathbf{z}_{n} = (r_n, t_n) \in \mathcal{X}$, and the corresponding output points given by $\mathbf{h} = [h_{1}, \ldots, h_{P}]^{\mathsf T} \in \mathbb{C}^P$, with $h_{n} = H_{r_{n}, t_{n}}$. Likewise, define the matrix of test points $\mathbf{Z}_{*} =[\mathbf{z}_{*, 1}^{\mathsf T}, \ldots, \mathbf{z}_{*, M}^{\mathsf T}]^{\mathsf T} \in \mathbb{N}^{M \times 2}$, with $\mathbf{z}_{*,m} = (r_m, t_m) \in \mathcal{X}_{*}$, and the corresponding output points given by $\mathbf{h}_{*} = [h_{*, 1}, \ldots, h_{*, M}]^{\mathsf T} \in \mathbb{C}^M$, with $h_{*,m} = H_{r_{m}, t_{m}}$. Our goal is to predict $\mathbf{h}_{*}$ from $\mathbf{h}$ via GPR.

Because the channel entries are complex, we model the real ($\mathrm{Re}$) and imaginary ($\mathrm{Im}$) parts as two different real-valued GPs that share the same covariance $k_{\boldsymbol{\theta}}(\cdot,\cdot)$. For each $\mathrm{Re}/\mathrm{Im}$ part, the joint real multivariate distribution is
\begin{equation}
\begin{bmatrix}
\mathbf{h}^{\mathrm{Re}/\mathrm{Im}} \\
\mathbf{h}_*^{\mathrm{Re}/\mathrm{Im}}
\end{bmatrix}
\sim \mathcal{N}\left(
\mathbf{0},
\begin{bmatrix}
\mathbf{K}+\sigma_\textrm{n}^2\mathbf{I}_P & \mathbf{K}_*^{\mathsf T} \\
\mathbf{K}_* & \mathbf{K}_{**}
\end{bmatrix}
\right),
\label{eq:joint_distribution}
\end{equation}
{with $\mathbf{K}=k_{\boldsymbol{\theta}}(\mathbf{Z},\mathbf{Z})\in\mathbb{R}^{P\times P}$,
$\mathbf{K}_*=k_{\boldsymbol{\theta}}(\mathbf{Z},\mathbf{Z}_*)\in\mathbb{R}^{P\times M}$, and
$\mathbf{K}_{**}=k_{\boldsymbol{\theta}}(\mathbf{Z}_*,\mathbf{Z}_*)\in\mathbb{R}^{M\times M}$.}
The posterior is given by
\begin{equation}
p(\mathbf{h}_*^{\mathrm{Re}/\mathrm{Im}}\mid \mathbf{h}^{\mathrm{Re}/\mathrm{Im}})
=\mathcal{N}\big(\boldsymbol{\mu}_*^{\mathrm{Re}/\mathrm{Im}},\boldsymbol{\Sigma}_*^{\mathrm{Re}/\mathrm{Im}}\big),
\label{eq:posterior_distribution}
\end{equation}
with
$
\boldsymbol{\mu}_*^{\mathrm{Re}/\mathrm{Im}}=\mathbf{K}_*^{\mathsf T}(\mathbf{K}+\sigma_\textrm{n}^2\mathbf{I}_P)^{-1}\mathbf{h}^{\mathrm{Re}/\mathrm{Im}}$ and $
\boldsymbol{\Sigma}_*^{\mathrm{Re}/\mathrm{Im}}=\mathbf{K}_{**}-\mathbf{K}_*^{\mathsf T}(\mathbf{K}+\sigma_\textrm{n}^2\mathbf{I}_P)^{-1}\mathbf{K}_*.
$

\textit{\textbf{Kernel selection for GPR:}}
The kernel function $k_{\boldsymbol{\theta}}(\cdot,\cdot)$ encodes prior beliefs about the spatial structure of $\mathbf{H}$. We evaluate our proposed framework considering three different kernels, namely, RBF, Mat\'ern, and RQ. These are chosen as they do not require an explicit channel model and span a wide range of smoothness and scale behaviors. The RBF kernel is
$k_{\textrm{RBF}}(\mathbf{z},\mathbf{z}') = \gamma_{\textrm{RBF}} \exp \bigl(-\tfrac{\|\mathbf{z}-\mathbf{z}'\|^2}{2\ell_{\textrm{RBF}}^2}\bigr)$, 
modeling smooth spatial variations via the lengthscale $\ell_{\textrm{RBF}}$ and scaling $\gamma_{\textrm{RBF}}$. The Mat\'ern kernel is
$k_{\textrm{Mat}}(\mathbf{z},\mathbf{z}') = \gamma_{\textrm{Mat}} \frac{2^{1-\nu}}{\Gamma(\nu)} \bigl(\tfrac{\sqrt{2\nu}\|\mathbf{z}-\mathbf{z}'\|}{\ell_{\textrm{Mat}}}\Bigr)^{\nu} \mathrm{K}_\nu \bigl(\tfrac{\sqrt{2\nu}\|\mathbf{z}-\mathbf{z}'\|}{\ell_{\textrm{Mat}}}\bigr)$,
controlling smoothness via $\nu$. Lastly, the RQ kernel is
$k_{\textrm{RQ}}(\mathbf{z},\mathbf{z}') = \gamma_{\textrm{RQ}} \bigl(1 + \tfrac{\|\mathbf{z}-\mathbf{z}'\|^2}{2\alpha \ell_{\textrm{RQ}}^2}\bigr)^{-\alpha}$,
generalizing RBF by multiple scales through $\alpha$.

In the Kronecker~\cite{kronecker2002} and Weichselberger~\cite{weichselberger2006} models, $\mathrm{vec}(\mathbf{H})$ is distributed as $\CN(\cdot,\cdot)$, which aligns naturally with a GP prior. To motivate the proposed GPR framework with RBF, Mat\'ern, or RQ kernels, Proposition~\ref{prop:blup} shows that the GP posterior mean equals the BLUP~\cite{BLUP}. Thus, optimality holds under second-order structure without requiring channel Gaussianity.

\begin{proposition}[GP posterior mean equals BLUP]\label{prop:blup}
Let $\mathbf u=\mathrm{vec}(\mathbf{H})\in\mathbb C^{N_{\textrm r}N_{\textrm t}}$ be a zero-mean, proper complex random vector with covariance $\mathbf{R_\mathrm{u}}\succeq\mathbf 0$ induced by any positive semidefinite kernel (e.g., RBF, Mat\'ern, or RQ). Let $\mathbf B\in\mathbb C^{P\times N_{\textrm r}N_{\textrm t}}$ select $P$ entries and consider the observation $\mathbf{H}=\mathbf B\mathbf u+\boldsymbol\varepsilon$, with $\boldsymbol\varepsilon\sim\mathcal{CN}(\mathbf 0,\sigma_{\textrm n}^2\mathbf I_P)$ and $\boldsymbol\varepsilon\indep\mathbf u$. Then, the BLUP of $\mathbf u$ from $\mathbf{H}$ under the second-order model $(\mathbf{R_\mathrm{u}},\sigma_{\textrm n}^2)$~is
\begin{equation}
\widehat{\mathbf u}_{\textrm{BLUP}}
= \mathbf{R_\mathrm{u}}\,\mathbf B^{\mathsf{H}}\big(\mathbf B\,\mathbf{R_\mathrm{u}}\,\mathbf B^{\mathsf{H}}+\sigma_{\textrm n}^2\mathbf I_P\big)^{-1}\mathbf{H}.
\label{eq:blup_map}
\end{equation}
Moreover, if $\mathbf u\sim\mathcal{CN}(\mathbf 0,\mathbf{R_\mathrm{u}})$ and 
$\boldsymbol\varepsilon\sim\mathcal{CN}(\mathbf 0,\sigma_{\textrm n}^2\mathbf I_P)$, 
then \eqref{eq:blup_map} coincides with the GP posterior mean estimator. 
Importantly, this equality is entirely determined by the second-order model 
$(\mathbf{R_\mathrm{u}},\sigma_{\textrm{n}}^2)$ and does not require $\mathbf u$ to be Gaussian.

\end{proposition}

\begin{IEEEproof}
Consider the linear predictors $\widehat{\mathbf u}=\mathbf L\mathbf{H}$, with $\mathbf L\in\mathbb C^{(N_{\textrm r}N_{\textrm t})\times P}$. Since we have $\mathbb E[\mathbf u]=\mathbf 0$ and $\mathbb E[\mathbf{H}]=\mathbf 0$, every linear predictor is unbiased. The mean squared error is $J(\mathbf L)=\mathbb E\|\mathbf u-\mathbf L\mathbf{H}\|_2^2
=\operatorname{tr}(\mathbf{R_\mathrm{u}})
-2\,\Re\{\operatorname{tr}(\mathbf L\,\mathbf R_{hu})\}
+\operatorname{tr}(\mathbf L\,\mathbf R_{hh}\,\mathbf L^{\mathsf{H}})$, with $\mathbf R_{hu}=\mathbb E[\mathbf{H}\mathbf u^{\mathsf{H}}]=\mathbf B\mathbf{R_\mathrm{u}}$ and
$\mathbf R_{hh}=\mathbb E[\mathbf{H}\mathbf{H}^{\mathsf{H}}]
=\mathbf B\mathbf{R_\mathrm{u}}\mathbf B^{\mathsf{H}}+\sigma_{\textrm n}^2\mathbf I_P$.
Using complex matrix calculus, $\partial J/\partial \mathbf L^\ast
=-\,\mathbf R_{hu}^{\mathsf{H}}+\mathbf L\,\mathbf R_{hh}$, so the unique minimizer (since $\mathbf R_{hh}\succ\mathbf 0$ when $\sigma_{\textrm n}^2>0$) is $\mathbf L^\star=\mathbf R_{hu}^{\mathsf{H}}\mathbf R_{hh}^{-1}
=\mathbf{R_\mathrm{u}}\,\mathbf B^{\mathsf{H}}\big(\mathbf B\mathbf{R_\mathrm{u}}\mathbf B^{\mathsf{H}}+\sigma_{\textrm n}^2\mathbf I_P\big)^{-1}$, which yields \eqref{eq:blup_map}. Under a Gaussian prior and Gaussian noise, the joint law of $(\mathbf u,\mathbf{H})$ is zero-mean with covariances $\mathbf{R_\mathrm{u}}$, $\mathbf B\mathbf{R_\mathrm{u}}$, and $\mathbf B\mathbf{R_\mathrm{u}}\mathbf B^{\mathsf{H}}+\sigma_{\textrm n}^2\mathbf I_P$. Finally, conditioning gives $\mathbb E[\mathbf u\mid \mathbf{H}]=\mathbf{R_\mathrm{u}}\,\mathbf B^{\mathsf{H}}\big(\mathbf B\mathbf{R_\mathrm{u}}\mathbf B^{\mathsf{H}}+\sigma_{\textrm n}^2\mathbf I_P\big)^{-1}\mathbf{H}$, which yields the BLUP.
\end{IEEEproof}

\textit{\textbf{Hyperparameter optimization:}}
Let $\boldsymbol{\theta}=[\gamma,\ell,\alpha,\nu]$ denote the kernel hyperparameters. For each $\mathrm{Re}/\mathrm{Im}$ part, the log marginal likelihood is $\log p(\mathbf{h}^{\mathrm{Re}/\mathrm{Im}}\mid\mathbf{Z},\boldsymbol{\theta})=-\frac{1}{2}(\mathbf{h}^{\mathrm{Re}/\mathrm{Im}})^{\mathsf T}(\mathbf{K}+\sigma_\textrm{n}^2\mathbf{I}_P)^{-1}\mathbf{h}^{\mathrm{Re}/\mathrm{Im}}-\frac{1}{2}\log\det(\mathbf{K}+\sigma_\textrm{n}^2\mathbf{I}_P)-\frac{P}{2}\log(2\pi)$. Its derivative with respect to each hyperparameter $\theta_i\in\{\gamma,\ell,\alpha,\nu\}$ is $\frac{\partial}{\partial\theta_i}\log p=\frac{1}{2}\mathrm{tr}[(\boldsymbol{\alpha}\boldsymbol{\alpha}^{\mathsf T}-(\mathbf{K}+\sigma_\textrm{n}^2\mathbf{I}_P)^{-1})\frac{\partial \mathbf{K}}{\partial \theta_i}]$, with $\boldsymbol{\alpha}=(\mathbf{K}+\sigma_\textrm{n}^2\mathbf{I}_P)^{-1}\mathbf{h}^{\mathrm{Re}/\mathrm{Im}}$. The optimal $\boldsymbol{\theta}^\star$ is obtained by maximizing $\log p(\mathbf{h}^{\mathrm{Re}}\mid\mathbf{Z},\boldsymbol{\theta})+\log p(\mathbf{h}^{\mathrm{Im}}\mid\mathbf{Z},\boldsymbol{\theta})$ using gradient-based methods~\cite{Jieao2025IEEE_TIT}.

\textit{\textbf{Complex channel reconstruction:}}
The $\mathrm{Re}$ and $\mathrm{Im}$ components of the channel are modeled by two GPs, $h_{\mathrm{Re}}(r,t)$ and $h_{\mathrm{Im}}(r,t)$, sharing the same kernel and hyperparameters. For a test location $\mathbf z_*=(r,t)$, the complex prediction is
$[\widehat{H}_{\textrm{GPR}}]_{r,t}
=\hat{h}_{\mathrm{Re}}(\mathbf z_*)+\jj\hat{h}_{\mathrm{Im}}(\mathbf z_*),$
where the posterior means $\hat{h}_{\mathrm{Re}}(\mathbf z_*)$ and $\hat{h}_{\mathrm{Im}}(\mathbf z_*)$ are obtained from~\eqref{eq:posterior_distribution}.
This preserves the shared second-order structure while avoiding cross-correlation assumptions between the $\mathrm{Re}$ and $\mathrm{Im}$ components.

\begin{figure*}[!t]
  \centering
  \begin{subfigure}{0.5\textwidth}
    \includegraphics[width=\linewidth,trim=0.01pt 0 0.01pt 0,clip]
    {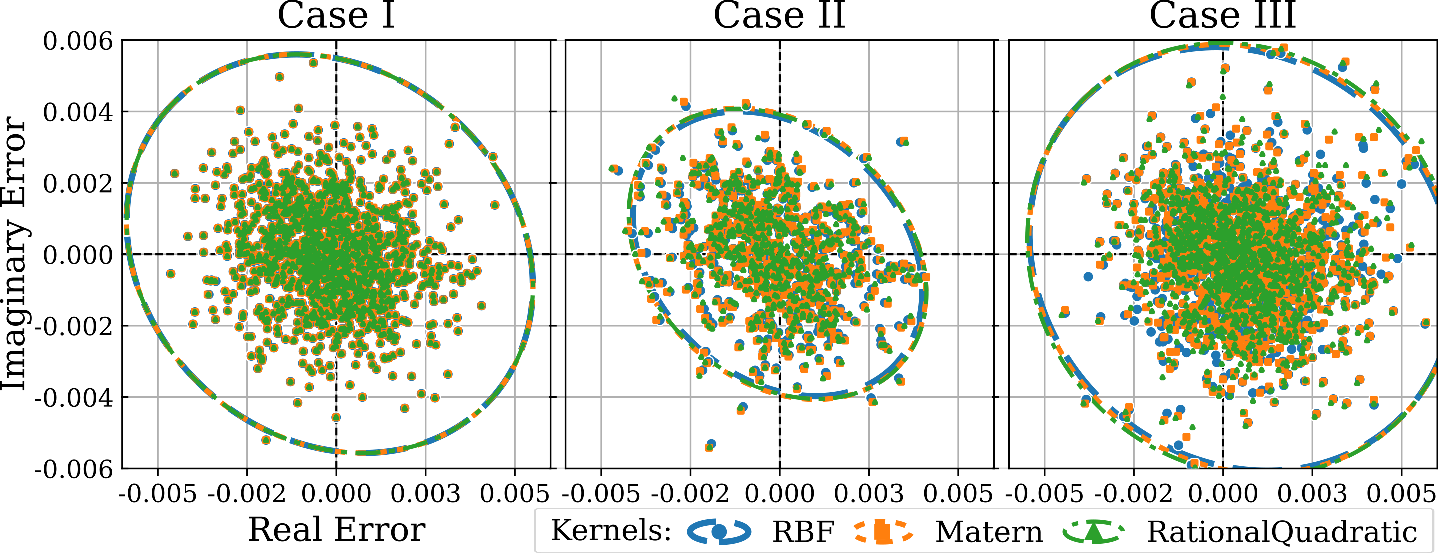}
    \caption{Kronecker model.}\label{fig:kron}
  \end{subfigure}\hfill
  \begin{subfigure}{0.5\textwidth}
    \includegraphics[width=\linewidth,trim=0.01pt 0 0.01pt 0,clip]
     {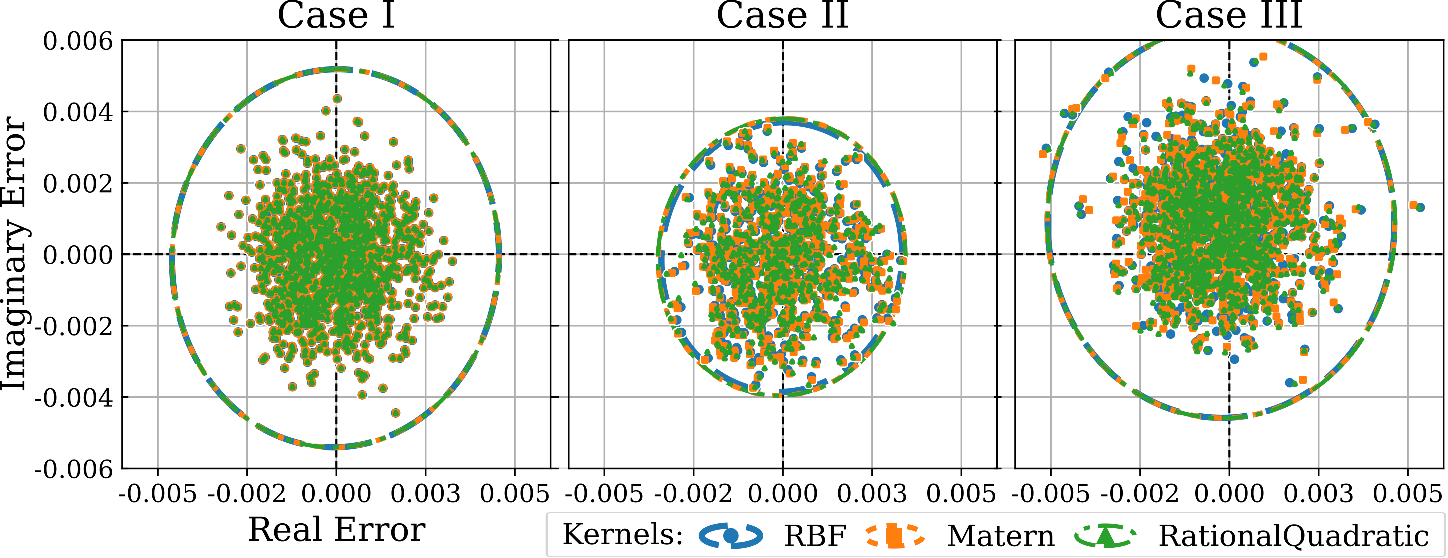}
    \caption{Weichselberger model.}\label{fig:weich}
  \end{subfigure}
  \caption{{Scatter plots of the prediction error illustrating the $\mathrm{Re}$ and $\mathrm{Im}$ error components for different training scenarios.}}
  \label{fig:ScatterError}
\end{figure*}

\begin{table}[tb]
\centering
\renewcommand{\arraystretch}{1.1}
\begingroup
\begin{tabular}{|l|c|c|}
\hline
\textbf{System Parameter} & \textbf{Symbol} & \textbf{Value} \\
\hline
\hline
\multicolumn{3}{|c|}{\textbf{Channels}} \\
\hline
Receive $\times$ transmit antennas & $N_{\textrm{r}}\times N_{\textrm{t}}$ & $36\times36$ \\
Element spacing & $d/\lambda$ & $0.5$ \\
Carrier frequency & $f_c$ & $28~\mathrm{GHz}$ \\
Angle spread & $\sigma_\phi$ & $\pi/6$ \\
\hline
\hline
\multicolumn{3}{|c|}{\textbf{Training (LS/MMSE baselines)}} \\
\hline
Pilot length (orthogonal) & $T$ & $T\ge N_{\textrm{t}}$, $\mathbf X\in\mathbb C^{N_{\textrm{t}}\times T}$ \\
SNR (unit power) & $\rho$ & $1/\sigma_\textrm{n}^2$ \\
\hline
\hline
\multicolumn{3}{|c|}{\textbf{GPR configuration}} \\
\hline
Kernel scale (init; bounds) & $\gamma$ & $1.0$; $[10^{-2},\,10^{2}]$ \\
RBF lengthscale (init; bounds) & $\ell_{\textrm{RBF}}$ & $0.5$; $[10^{-2},10]$ \\
Mat\'ern lengthscale & $\ell_{\textrm{Mat}}$ & $0.5$; $[10^{-2},10]$ \\
Mat\'ern smoothness & $\nu$ & $1.5$ (fixed) \\
RQ lengthscale & $\ell_{\textrm{RQ}}$ & $0.5$; $[10^{-2},5]$ \\
RQ shape & $\alpha$ & $0.5$; $[10^{-1},5]$ \\
\hline
\end{tabular}
\endgroup
\caption{{{Simulation parameters and hyperparameters.}}}
\label{tab:params}
\end{table}

\section{Simulation Results and Discussion}
We simulate the proposed schemes on a $36\times36$ link, under both Kronecker~\cite{kronecker2002} and Weichselberger~\cite{weichselberger2006} channel models. Unless stated otherwise, the simulation parameters summarized in Table~\ref{tab:params} are adopted. We consider the three pilot probing schemes of Fig.~\ref{fig:System_Model}. Specifically: Case~I observes a single column of $\mathbf{H}$ (i.e., $N_{\textrm{r}}$ entries) and predicts the remaining $N_{\textrm{t}}-1$ columns via GPR; Case~II observes half of the columns (i.e., approximately $N_{\textrm{r}}N_{\textrm{t}}/2$ entries) and predicts the unobserved half; Case~III observes only the diagonal entries ($\min(N_{\textrm{r}},N_{\textrm{t}})$) and predicts all the off-diagonal entries from the correlation structure. Note that Case~I and Case~II lead to a $\frac{(N_\textrm{t}-1)}{N_\textrm{t}}$ and $50\%$ pilot-reduction, respectively.

To assess the prediction accuracy, uncertainty calibration, and communication impact, we employ three metrics. 
First, we compute the complex entrywise prediction error $\epsilon_{ij}= H_{ij}-\widehat{H}_{ij}$, visualized via scatter plots of $\mathrm{Re}(\epsilon_{ij})$ versus $\mathrm{Im}(\epsilon_{ij})$ with 95\% error ellipses.
Second, to quantify how well each estimator preserves the channel structure relevant for data transmission, we evaluate the SE of a multi-stream MIMO link under a linear receiver designed from the estimated channel. For a given channel estimate $\widehat{\mathbf{H}}\in\mathbb{C}^{N_{\textrm{r}}\times N_{\textrm{t}}}$, we adopt the linear MMSE detector $\mathbf{W}(\widehat{\mathbf{H}})= \big( \widehat{\mathbf{H}}\widehat{\mathbf{H}}^{\mathsf{H}} + \frac{N_{\textrm{t}}}{\rho}\,\mathbf{I}_{N_{\textrm{r}}} \big)^{-1}\widehat{\mathbf{H}}=[\mathbf{w}_1,\ldots,\mathbf{w}_{N_{\textrm{t}}}]\in\mathbb{C}^{N_{\textrm{r}}\times N_{\textrm{t}}}$. Let $\mathbf{h}_{k}$ and $\mathbf{w}_{k}$ denote the $k$th columns of the true channel $\mathbf{H}$ and the detector $\mathbf{W}$, respectively. The post-equalization signal-to-interference-plus-noise ratio (SINR) of stream $k$ is then

\begin{equation}
\mathrm{SINR}_{k}(\widehat{\mathbf{H}})
= \frac{\bigl|\mathbf{w}_{k}^{\mathsf{H}}\mathbf{h}_{k}\bigr|^2}
{\sum_{j\neq k}\bigl|\mathbf{w}_{k}^{\mathsf{H}}\mathbf{h}_j\bigr|^2 
+ \frac{N_{\textrm{t}}}{\rho}\,\|\mathbf{w}_{k}\|^2},
\end{equation}
where the true channel $\mathbf{H}$ is always used inside the SINR expression, while the estimate $\widehat{\mathbf{H}}$ affects performance only through the detector $\mathbf{W}(\widehat{\mathbf{H}})$. The corresponding SE is $\mathrm{SE}(\widehat{\mathbf{H}})
= \sum_{k=1}^{N_{\textrm{t}}} \log_2\bigl(1+\mathrm{SINR}_{k}(\widehat{\mathbf{H}})\bigr)$, which we evaluate for the true channel $\mathbf{H}$, the GPR-predicted channel $\widehat{\mathbf{H}}_{\textrm{GPR}}$, and the LS and MMSE estimates $\widehat{\mathbf{H}}_{\textrm{LS}}$ and $\widehat{\mathbf{H}}_{\textrm{MMSE}}$, respectively. Only for the LS and MMSE baselines, we adopt full-array orthogonal training with length $T \ge N_{\textrm{t}}= n_{\textrm{t}}$ and pilot matrix $\mathbf{S} \in \mathbb{C}^{N_{\textrm{t}}\times T}$. The LS estimate equals
$
\widehat{\mathbf{H}}_{\textrm{LS}} = \mathbf{Y}\mathbf{S}^{\mathsf{H}} (\mathbf{S}\mathbf{S}^{\mathsf{H}})^{-1}.
$
Let $\mathbf{y}= \mathrm{vec}(\mathbf{Y})$. The MMSE estimator equals
$
\widehat{\mathbf{u}}_{\textrm{MMSE}} = \mathbf{R}_{\textrm{H}}\mathbf{A}^{\mathsf{H}}(\mathbf{A}\mathbf{R}_{\textrm{H}}\mathbf{A}^{\mathsf{H}}+\sigma_\textrm{n}^2\mathbf{I}_{T N_{\textrm{r}}})^{-1}\mathbf{y}$, with
$
\mathbf{A}= \mathbf{S}^{\mathsf T}\otimes \mathbf{I}_{N_{\textrm{r}}}
$
and $\widehat{\mathbf{H}}_{\textrm{MMSE}}= \mathrm{unvec}(\widehat{\mathbf{u}}_{\textrm{MMSE}})$. Finally, we evaluate the uncertainty calibration via the empirical coverage of marginal 95\% credible intervals $\widehat{H}_{ij} \pm 1.96\sigma_{ij}$, i.e., the fraction of the true $H_{ij}$ that falls within these intervals derived from the GPR posterior variance.

\paragraph{Prediction error with 95\% confidence tolerance} Fig.~\ref{fig:ScatterError} shows, for each kernel, the joint scatter plots of the $\mathrm{Re}$ and $\mathrm{Im}$ components of the per-entry prediction error $\epsilon_{ij}$ under the three pilot-probing schemes (Cases~I-III) for both the Kronecker (Fig.~\ref{fig:kron}) and Weichselberger ( Fig.~\ref{fig:weich}) channel models. Each scatter cloud is overlaid with a 95\% confidence ellipse computed from the empirical covariance of the error pairs, representing the region where 95\% of the bivariate errors are expected to lie. The ellipse area reflects the total error variance, with its orientation and eccentricity indicating the correlation and anisotropy between the $\mathrm{Re}$ and $\mathrm{Im}$~parts.

In Case~I, the prediction errors are largest across all kernels, the scatter is widely spread, and the confidence ellipses are expansive. In Case~II, activating 50\% of the transmit antennas produces a clear contraction of the error clouds toward the origin. The corresponding ellipses shrink in area and become more circular, indicating improved prediction accuracy due to the larger number of observed entries. The Kronecker model retains more elongated ellipses (especially in Cases~I and~III), signifying persistent anisotropy, and stronger directional correlation. In contrast, the Weichselberger model exhibits nearly circular ellipses, reflecting its richer spatial coupling that balances the error variance more evenly between components.

The prediction errors in Case~III are higher than in Case~II but lower than in Case~I. Although Cases~I and~III use the same number of observations, diagonal sampling provides two-dimensional coverage and better anchors the channel energy/principal subspace that GPR exploits, albeit at the cost of higher pilot overhead, whereas a single column lacks cross-row diversity. Case~III is considered to highlight the important fact that the performance of GPR not only depends on the number of observations, but also on how those observations are sampled. Case~II performs best because its equispaced half-column design offers both more pilots and richer cross-column information for GPR. The Kronecker model continues to exhibit anisotropic errors, while the Weichselberger model maintains balanced, isotropic error distributions. These results support three main conclusions: (i) increasing the number of directly observed channel entries consistently improves the prediction accuracy; (ii) the Weichselberger model’s richer spatial correlation yields more balanced error statistics than the separable Kronecker model; and (iii) all the considered kernels deliver comparable and well-calibrated predictions across the different pilot-probing schemes. Subsequent analyses are presented exclusively using the Weichselberger channel due to its greater physical realism~\cite{weichselberger2006}.

\begin{figure}
    \centering
    \includegraphics[width=0.9\linewidth]{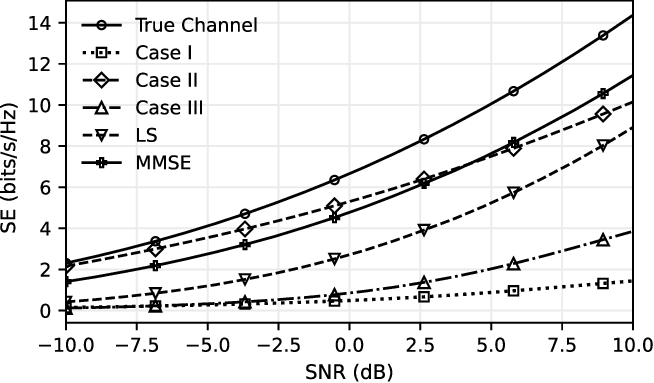}
    \caption{SE versus SNR for Cases~I--III with the Mat\'ern kernel, against LS/MMSE and the true channel.}
    \label{fig:MI}
\end{figure}

\paragraph{Spectral efficiency}
Fig.~\ref{fig:MI} shows the SE of the GPR-predicted channels for Cases~I--III with the Mat\'ern kernel, compared against the true Weichselberger channel and the full-pilot LS and MMSE baselines. Case~II consistently achieves the highest SE among the pilot-saving schemes across the whole SNR range, and clearly outperforms LS. It also exceeds MMSE at low-to-moderate SNR and remains close to it at higher SNR. In relative terms, Case~II preserves about $94\%$ of the perfect-CSI SE at $\mathrm{SNR}=-10$\,dB and about $71\%$ at $\mathrm{SNR}=10$\,dB, while using only $50\%$ of the pilots. This follows from probing half of the transmit columns, which yields dense and well-distributed observations over the antenna grid; the GPR posterior then acts as a data-adaptive spatial regularizer that learns the correlation structure, suppresses noise, and preserves the dominant modes that drive SE under the adopted linear receiver.

The MMSE baseline provides the strongest conventional performance because it is built from the true second-order channel statistics derived from the joint eigenstructure of the Weichselberger model, but it requires perfect knowledge of these statistics and full-array pilots. In contrast, Case~II attains comparable SE without assuming a parametric channel model and with only half of the pilot overhead. On the other hand, LS is highly noise-sensitive and does not exploit spatial correlation, which results in substantially lower SE. At $0$\,dB, for example, LS attains only $39.52\%$ of the perfect-CSI SE, compared with $80.24\%$ for Case~II and $71.19\%$ for MMSE, as summarized in Table~\ref{tab:pilot_MI_GPR} for the RQ kernel.

Cases~I and~III use the same number of observations but different sampling geometries. Case~III consistently achieves higher SE than Case~I, highlighting the role of spatial coverage: diagonal sampling provides joint transmit–receive supervision, whereas column sampling excites only a single transmit dimension. However, in both cases the available supervision remains insufficient to identify the full joint channel structure required for spatial multiplexing, leading to poor stream separability and residual multi-stream interference.

For all the three cases, the RBF, Mat\'ern, and RQ kernels produce very similar SE curves against the SNR, indicating robustness to the kernel choice under the considered sampling geometries. After hyperparameter optimization, the learned lengthscales become large relative to the array aperture, so the induced covariance structures are nearly indistinguishable and impose similar spatial priors. Consequently, the SE trends observed with the Mat\'ern kernel in Fig.~\ref{fig:MI} are consistent with those reported for the RQ kernel in Table~\ref{tab:pilot_MI_GPR}.

\begin{table}[tb]
\centering
\renewcommand{\arraystretch}{1.1}
\begingroup
\begin{tabular}{|l|c|c|c|c|c|}
\hline
\textbf{Metric} & \textbf{Case~I} & \textbf{Case~II} & \textbf{Case~III} & \textbf{LS} & \textbf{MMSE} \\
\hline
\hline
Pilot length $T$ & $1$ & $18$ & $36$ & $36$ & $36$ \\
Pilot saving [\%] & \textbf{$97.22$} & $50.00$ & $0.00$ & $0.00$ & $0.00$ \\
Relative SE [\%] & $7.23$ & \textbf{$80.24$} & $4.05$ & $39.52$ & $71.19$ \\
\hline
\end{tabular}
\endgroup
\caption{Pilot saving and relative SE at 0\,dB relative to the true channel for GPR-RQ, LS, and MMSE.}
\label{tab:pilot_MI_GPR}
\end{table}

\begin{figure} 
    \centering
    \includegraphics[width=0.8\linewidth]{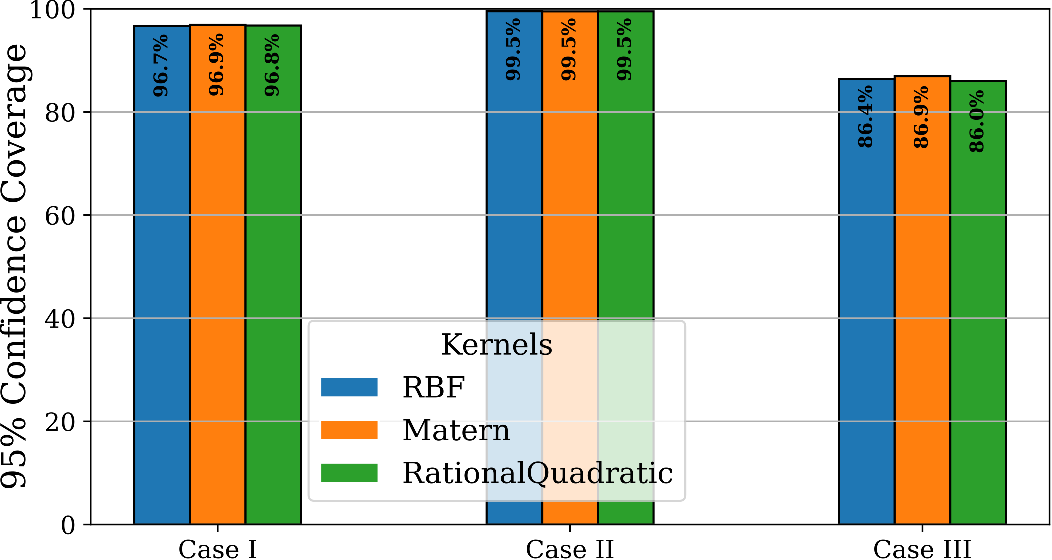} 
    \caption{{Empirical 95\% credible-interval coverage rates.}}
    \label{fig:predictionConfidence}
\end{figure} 

\paragraph{Uncertainty calibration}
To evaluate uncertainty quantification, Fig.~\ref{fig:predictionConfidence} reports the empirical coverage of the marginal 95\% credible intervals, i.e., the fraction of true entries that fall within $\widehat{H}_{ij}\pm 1.96\sigma_{ij}$ from the GPR posterior. A well-calibrated model attains coverage close to the nominal 95\% (where higher indicates conservative uncertainty and lower indicates overconfidence). 
The RBF, Mat\'ern, and RQ kernels generally exhibit nearly identical coverage. The learned hyperparameters drive $\ell$ to large values relative to the array aperture, making these isotropic kernels induce almost constant correlation over the observed antenna grid. As a result, their posterior means and variances, and thus their marginal credible intervals, are effectively the same. This behavior is reinforced by the sampling geometries in Cases~I/III (e.g., predominantly collinear/diagonal observations), under which minor differences in kernel shape become negligible.    

\section{Conclusion}
This paper proposed a GPR-based framework for MIMO channel estimation that reconstructs full CSI from a small subset of pilot observations. By employing RBF, Mat\'ern, and RQ kernels, the framework models smooth and multi-scale spatial correlations without requiring prior channel statistics. Simulations on Kronecker and Weichselberger channels show that, in Case~II, the method reduces pilots by $50\%$ while preserving over $80.24\%$ of the true-channel SE, outperforming LS and MMSE baselines. These results indicate that GPR enables accurate, uncertainty-aware channel reconstruction with substantial training-overhead savings for next-generation MIMO systems. Future work will design kernel structures that more precisely encode channel statistics to further improve estimation performance.
\bibliographystyle{IEEEtraN_renamed}
\bibliography{bibliography}
\end{document}